\documentclass{article}
\usepackage{waspaa21,amsmath,amssymb,graphicx,url,times}
\usepackage{xcolor}
\usepackage{float}

\usepackage{mathrsfs}
\newcommand{\ve}[1]{\textbf{#1}}		
\DeclareMathOperator{\E}{\mathbb{E}}
\DeclareMathOperator{\softmax}{softmax}

\title{Adversarial Auto-Encoding for Packet Loss Concealment}

\name{Santiago Pascual, Joan Serr\`a, Jordi Pons}
\address{Dolby Laboratories\\ santiago.pascual@dolby.com\\  }

\begin{document}

\ninept
\maketitle

\begin{sloppy}

\begin{abstract}
Communication technologies like voice over IP operate under constrained real-time conditions, with voice packets being subject to delays and losses from the network. In such cases, the packet loss concealment (PLC) algorithm reconstructs missing frames until a new real packet is received. Recently, autoregressive deep neural networks have been shown to surpass the quality of signal processing methods for PLC, specially for long-term predictions beyond 60\,ms. In this work, we propose a non-autoregressive adversarial auto-encoder, named PLAAE, to perform real-time PLC in the waveform domain. PLAAE has a causal convolutional structure, and it learns in an auto-encoder fashion to reconstruct signals with gaps, with the help of an adversarial loss. During inference, it is able to predict smooth and coherent continuations of such gaps in a single feed-forward step, as opposed to autoregressive models. Our evaluation highlights the superiority of PLAAE over two classic PLCs and two deep autoregressive models in terms of spectral and intonation reconstruction, perceptual quality, and intelligibility.
\end{abstract}

\begin{keywords}
packet loss concealment, adversarial networks, auto-encoder, speech enhancement.
\end{keywords}

\section{Introduction}
\label{sec:intro}

Voice over IP communications are constrained to work in real-time conditions, so that they work fluidly and users do not feel that there are interruptions during the conversation. In order to fulfil these conditions, packets must be served quickly and continuously to the decoder so that the voice signal can be reconstructed and played. Nonetheless, several disruptions can happen in a communications network, and voice packets can get severely delayed or lost, disrupting the perception quality for the receiving user~\cite{takahashi2004perceptual}. The receiver device relies on the so-called jitter buffer to reorder and accumulate packets before decoding. When packets arrive too late such that a gap would be perceived, a packet loss concealment (PLC) algorithm is applied to reconstruct the missing packets, hence allowing the jitter buffer to continue serving packets. Basic PLC techniques may fill the gaps with zeros to substitute lost fragments, repeat preceding frames, or use some form of interpolation between them~\cite{mohamed2020deep}. 

Machine learning methods have also been used for PLC, upon the principle of predicting future acoustic frames given a sequence of past frames. Early statistical modeling approaches made use of Gaussian mixture~\cite{lindblom2000model} and hidden Markov~\cite{rodbro2006hidden} models to predict future acoustic parameters. Deep neural networks were also proposed to solve this task in the form of autoregressive fully-connected networks~\cite{lee2015packet}, outperforming previous statistical approaches in synthesis quality and robustness against higher loss rates. 
Also recurrent networks like LSTMs have been used as regression models of waveform samples to perform PLC in a real-time setup based on online learning~\cite{lotfidereshgi2018speech}. Recurrent networks are known to be very competitive for sequential generative modeling, and specially for speech synthesis. WaveRNN proves this, with a recurrent structure that is designed to be compact and to run at high speed~\cite{kalchbrenner2018efficient}. Despite its compact design, it yields high quality synthetic speech, comparable to that of much bigger and inefficient models like WaveNet~\cite{oord2016wavenet}. Due to these properties, a WaveRNN variant named WaveNetEQ~\cite{waveneteq} was proposed to perform real-time PLC, comparing favourably against a classic counterpart, specially for long-term losses beyond 60\,ms. 
Nonetheless, due to its generative capacity, special tuning of the sampling process must be considered to generate a future that matches the real signal coming after the lost packets. In WaveNetEQ, this is controlled with a fade out of the generation after 120\,ms, to ensure that it generates silence~\cite{waveneteq}. 

In contrast to autoregressive models, generative adversarial networks (GANs) generate speech in a parallel fashion and with compelling quality~\cite{binkowski2019high}. 
Importantly, the adversarial learning strategy can be adopted outside the generative framework. This way, the discriminator can be seen as a learnable loss function of what we want to achieve through examples of fake and real results~\cite{isola2017image} (note that we can even train regression models in an adversarial fashion to boost their prediction quality). GANs and deep auto-encoders have been previously used for speech reconstruction in the context of audio inpainting~\cite{chang2019deep, marafioti2019context, shi2019speech}, which is similar to PLC but without real-time restrictions. Therefore, these models can be slow, relatively complex, and can exploit future information coming after the lost packets, which is something that is not compatible with PLC. 

In this work, we propose the packet loss adversarial auto-encoder (PLAAE), a non-autoregressive model for PLC. The entire encoder-decoder structure is causal, to make it compatible with the real-time reconstruction task, and its inference is parallelized across the many packets that need to be recovered. Additionally, PLAAE yields waveform samples, which eliminates the need for acoustic feature inversion after reconstruction. 
PLAAE learns with both supervised and adversarial multi-resolution losses, which make reconstructions sound natural and coherent within their context.

\section{Packet Loss Adversarial Auto-Encoder}
\label{sec:plaae}

The architecture of PLAAE is shown in Fig.~\ref{fig:plaae}. The model ingests a spectral representation of the zero-filled lossy waveform and generates the reconstructed waveform samples. The reason for this asymmetric construction is twofold. Firstly, we boost the efficiency of training and inference with a compact spectral representation in the input, with a much lower sampling rate than the raw signal. Secondly, we boost the synthesis quality by directly generating the waveform, which can be achieved with recent training techniques like multi-resolution spectral and adversarial losses~\cite{kumar2019melgan, yang2021multi}. The input lossy signal is zero-filled for the full extension of the gap, and the model predicts the filling samples all in parallel. 

PLAAE takes the form of an auto-encoder, with both encoder and decoder being causal and fully convolutional. 
The encoder is composed of $N$ causal convolutional blocks, preceded by a layer normalization block~\cite{ba2016layer} and followed by a ReLU activation. Each convolutional layer has a dilation factor $d$ so that its receptive field spans exponentially with depth~\cite{oord2016wavenet}. Our current implementation features $N=5$~blocks and the dilation factor is $d = 3^{N}$. The encoder output is finally projected through a linear layer (1$\times$1 convolution) to obtain embeddings that get injected into the decoder. 

The decoder is composed of $M$ blocks of non-overlapping transposed convolutions (i.e.,~causal, since stride and kernel size are equal~\cite{pons2020upsampling}) and causal residual blocks.  The residual blocks feature the same type of convolutions as in the encoder, as well as a linear layer that projects the features back to the residual dimensionality to aggregate the features (Fig.~\ref{fig:plaae}, left). The convolution in the decoder's residual block also features a dilation determined by the factor $d=3^L$. During our experiments, we noticed that increasing the depth of the $m$-th decoder block (i.e.,~increasing its receptive field) enhanced the quality of the model output. This was also observed in other works like the multi-band MelGAN~\cite{yang2021multi}. Finally, the output of the decoder is linearly projected into a single channel to yield a mono signal, with a $\tanh$ limiting the output range to $[-1,1]$. All learnable layers in the decoder are weight-normalized to stabilize the adversarial training~\cite{kumar2019melgan}.

\begin{figure}[!t]
    \centering
    \includegraphics[width=0.95\linewidth]{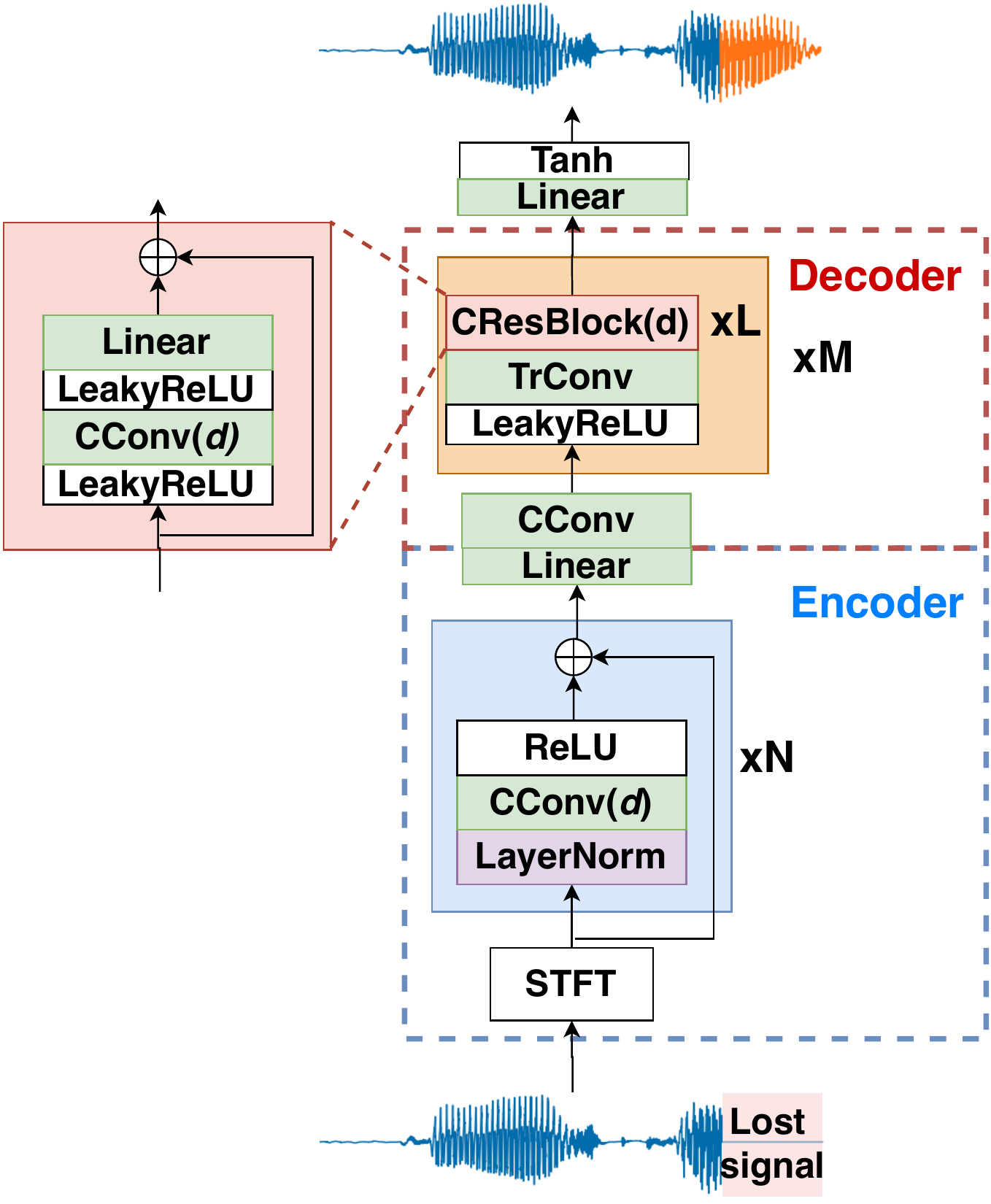}
    \vspace{-2mm}
    \caption{PLAAE architecture. Prefix ``C'' stands for causal, and ``Tr'' for transposed.}
    \label{fig:plaae}
\end{figure}

To train PLAAE, we employ three discriminators with equal architectures, but each processes the input signal at a different sampling rate: 16, 4, and 1\,kHz. The re-samplings take the form of average pooling layers, and their architecture follows the same deep convolutional structure from MelGAN with weight-normalized layers (as in the decoder). In this work, we rely on the least-squares GAN formulation~\cite{mao2017least}, 
\begin{equation}
\begin{aligned}
\underset{D}\min~V(D) & = \frac{1}{2}\E_{\ve{x}}\left[(D(\ve{x}) - 1)^{2}\right] + \frac{1}{2}\E_{\tilde{\ve{x}}}\left[D(G(\tilde{\ve{x}}))^{2}\right],
\end{aligned}
\end{equation}
where $\ve{x}$ is a non-lossy reference signal and $\tilde{\ve{x}}$ is a lossy one. On the other hand, PLAAE is trained with the loss
\begin{equation}
\begin{aligned}
\underset{G}\min~V(G) & = \E_{\tilde{\ve{x}}}\left[(D(G(\tilde{\ve{x}})) - 1)^{2}\right] + \\
& + \alpha \E_{\ve{x}, \tilde{\ve{x}}}\left[\frac{1}{R}\sum_{r=1}^{R} L_{\text{SC}}^{r}(\ve{x},\tilde{\ve{x}}) + L_{\text{MAG}}^r(\ve{x}, \tilde{\ve{x}})\right],
\end{aligned}
\end{equation}
where the first term is the adversarial one and the second term is a multi-resolution STFT (multi-STFT) loss~\cite{arik2018fast} scaled by a parameter $\alpha$. This loss is composed of a spectral convergence term ($L_{\text{SC}}$) and a $\log$-scale magnitude loss ($L_{\text{MAG}})$ for different $R$ spectral resolutions, which are obtained with the same STFT configurations as in~\cite{yang2021multi}. The multi-STFT boosts the synthesis quality and training stability in this waveform reconstruction scenario. Nevertheless, we noticed that the adversarial term is crucial to reproduce all the details in speech that are missed by the spectral loss, the most obvious one being a coherent phase. The two joint losses yielded high quality outputs, and we noticed that setting $\alpha=1$ worked well throughout different experiments.

Note that, with PLAAE, we prescind of the generative modeling part of GANs (i.e.,~there is no random input source). Instead, we have a regressor model that predicts always the same future trajectory given a past signal, without any inherent sampling that yields a randomization of the outcomes. During the development of PLAAE, we considered the usage of random latent variables, but this made the model predictions less coherent with the past and future signals that had to be cross-faded with the generation. Therefore, besides making PLAAE fully deterministic, we found that removing the generative sampling reduced the content and speaker variability in model's predictions.

\section{Experimental Setup}
\label{sec:exp_setup}

The lossy input signal is transformed into a mel spectrogram of 80~bins with 
an STFT analysis of $N=1024$~points, a Hann window of 20\,ms, and 50\% overlap. An additional channel with a binary flag is added to the acoustic features to signal whether the current silence is real or corresponds to lost packets. Note that this information is available at concealment time. PLAAE is then trained to reconstruct these features as a clean waveform and, due to its causal architecture, the task becomes directly available for the PLC application. Therefore, during inference, we can inject lossy signals by inserting zeros and the activated flag to denote that there is signal to be reconstructed, and the model will look into the available past through its receptive field to perform the reconstruction. The prediction of all future samples happens in parallel, and any past packet loss is not filled with predictions to proceed inferring. Instead, they remain in the past and the model must learn to reconstruct with missing components in the past, what forces it to be more resilient to consecutive lost packets. The amount of samples to predict for a packet loss is $T$, which corresponds to $P = \frac{T}{L}$ packets, being $L = 320$~samples at 16\,kHz sampling rate (i.e.,~a packet is 20\,ms long). PLAAE is trained with mini-batches of 48 1-second segments during 1.5\,M iterations, keeping track of the validation multi-STFT loss to approach convergence. The Adam optimizer is used for both PLAAE and the discriminators, with a learning rate of $3\cdot10^{-4}$ and betas of 0.5 and 0.9.

\subsection{Baselines}
\label{sec:baselines}
To assess the performance of PLAAE for PLC, we compare the quality of its reconstructions against a mixture of classic and deep learning baselines. First, we consider a zero-filling PLC, which establishes a reconstruction lower bound (reconstructing nothing). Secondly, we consider the use of the G722.1 ITU-T standard codec, with its inherent PLC~\cite{recommendation2005722}. Since the concealment outcome must blend smoothly with the original signal, the inclusion of G722.1 inputs makes us additionally evaluate the performance of PLAAE using a coded input signal. Thirdly, we consider WebRTC's PLC~\cite{webrtc}, named NetEQ, which works upon wideband PCM data. Therefore, we also include a signal processing baseline that operates upon non-codec speech. NetEQ uses signal processing methods to analyze the speech and produce smooth continuations that work well for short-term packet losses (e.g.,~20\,ms), but sounds robotic and unnatural for longer segments~\cite{waveneteq}.

As deep learning baselines we first consider an early 
auto-regressive DNN for PLC~\cite{lee2015packet}, where a frame of spectral magnitude and phase is predicted conditioned on a window of past frames. Our implementation follows the design of two separate DNN modules, the former processing magnitude and the latter processing phase, as it proved to be beneficial in prior implementations~\cite{lee2015packet}. The spectral frames are extracted from an STFT analysis equal to the one used for PLAAE. In our implementation, we used an L1 regression for the log-power spectrum and the instantaneous phase as proposed in recent neural vocoder works~\cite{arik2018fast}. We notice that instantaneous phase prediction converged faster and yielded better reconstructions than a direct regression on the raw phase. This model is trained to predict the next frame from the previous 11 packets for 1\,M steps using mini-batches of 128 and a learning rate of $10^{-3}$, until validation convergence. The last deep learning baseline we consider is based on a WaveRNN generative model. 
This model is trained to reconstruct the immediate next sample. It follows the basic structure of WaveNetEQ~\cite{waveneteq}, with our interpretation of an encoder to process the past signal that fits the model requirements. 
This encoder is composed of two fully connected residual blocks, a two-layer bidirectional LSTM and a linear projection. Finally, a statistical pooling layer through time is applied to extract the mean activations and their standard deviation, which are concatenated into a single vector. This is finally concatenated to the previously predicted sample to condition the WaveRNN. 
This model is trained for 1\,M iterations until validation convergence.

To smooth the insertion of predictions, we use linear cross-fading transitions of 25\% of a packet length, similar to other works~\cite{lee2015packet}. Models generate two extra frames to perform cross-fading. For both DNN and PLAAE, we match the prediction phase with the last real frame via maximum correlation. In the WaveRNN, we only apply a fade-out transition since its sample-wise process naturally transitions from past sample to reconstructed sample.

\subsection{Evaluation}
\label{sec:evaluation}

We measure the amount of signal distortion as gap length increases. 
To this end, we employ typical speech synthesis objective measurements that correlate with ground-truth speaker identity, spoken content, and intonation. In particular, we compute mel cepstral distortion (MCD; in dBs)~\cite{kubichek1993mel}, the root mean squared error of estimated F0 contours (F0), and voiced/unvoiced frame errors (UV)~\cite{pascual2020efficient}. We report MCD for every 10\,ms and averaged F0 and UV (estimated and interpolated pitch contours do not vary rapidly for the considered time span in our non-expressive data set, see below). 

In addition to signal distortion, we also evaluate the perceptual difference between concealed test utterances and their non-lossy counterparts. This emulates a realistic PLC scenario, where multiple packet losses can occur and the PLC algorithm operates several times per utterance. To that end we use SESQA~\cite{serra2020sesqa}, which is a semi-supervised model that correlates with human judgments of speech quality. Among other degradations, SESQA has been trained with programatically-generated packet losses and a diversity of codecs, hence we expect it to highlight differences in the perceived reconstruction quality to a reasonable extent. 
Finally, we also assess the resulting intelligibility after concealment with a pre-trained acoustic model for automatic speech recognition using the character error rate (CER, \%). For that, we use the DeepSpeech2-V2 release model, which is pre-trained on the LibriSpeech dataset~\cite{panayotov2015librispeech}, taken from an open source implementation\footnote{https://github.com/SeanNaren/deepspeech.pytorch}.

\subsection{Dataset}

We run our experiments on the VCTK corpus~\cite{vctk}, which contains tens of hours of neutral speech uttered by 109~speakers. We first pre-processed the corpus with a voice activity detector to filter out silences longer than 100\,ms at the beginning and end of files and resampled it to 16\,kHz. Secondly, we split the corpus into train, validation, and test subsets stratifying speaker identity, so they contain 99, 4, and 6 speakers, respectively (note that these splits are gender-balanced as well). We additionally filter out test utterances that are shorter than 2\,s due to our study on long packet losses, where there may not be enough signal history prior to a large gap. After this process, we obtain 23\,h of training speech, 1\,h for validation, and 1\,h for test. Furthermore, apart from the PCM we also include codec conditions during training by means of an internal speech communications codec and G722.1, with a maximum bitrate of 24\,kbps.

Once we have the test split, we create two subsets to be used in the two evaluation scenarios described in section~\ref{sec:evaluation}. First, we randomly extract a segment of one second per test utterance as past example, and its subsequent 120\,ms are taken as future ground-truth. This results in approximately 1300 segments to carry out signal distortion measurements. Secondly, we inject packet losses to the full test set with lengths of 20, 40, 60, and 120\,ms and a packet drop probability of 0.1 for non-contiguous packets. With that we obtain four copies of the test set, each with a certain packet loss length. 

\section{Results}
\label{sec:results}

\begin{figure*}[t!]
    \centering
    \includegraphics[width=0.48\textwidth]{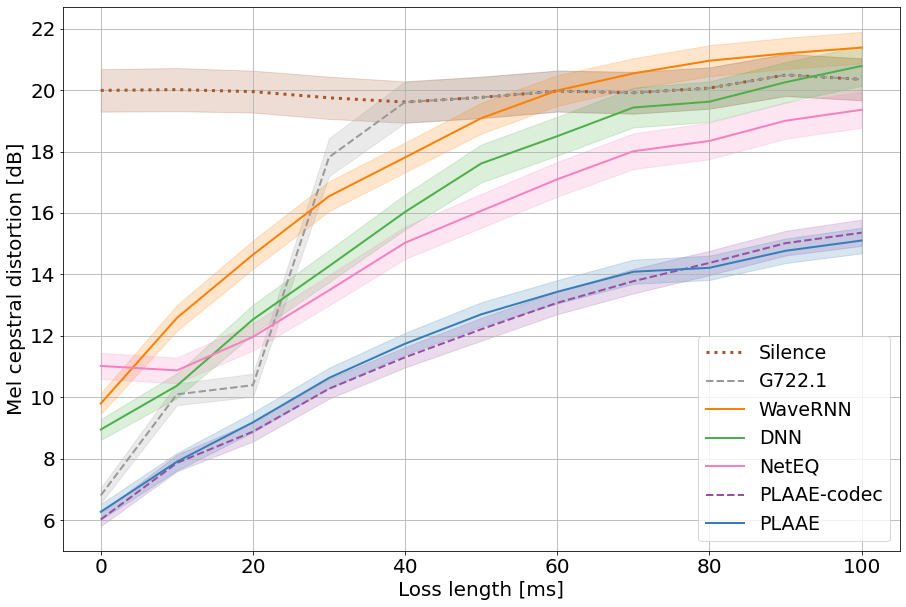} \hfill
    \includegraphics[width=0.48\textwidth]{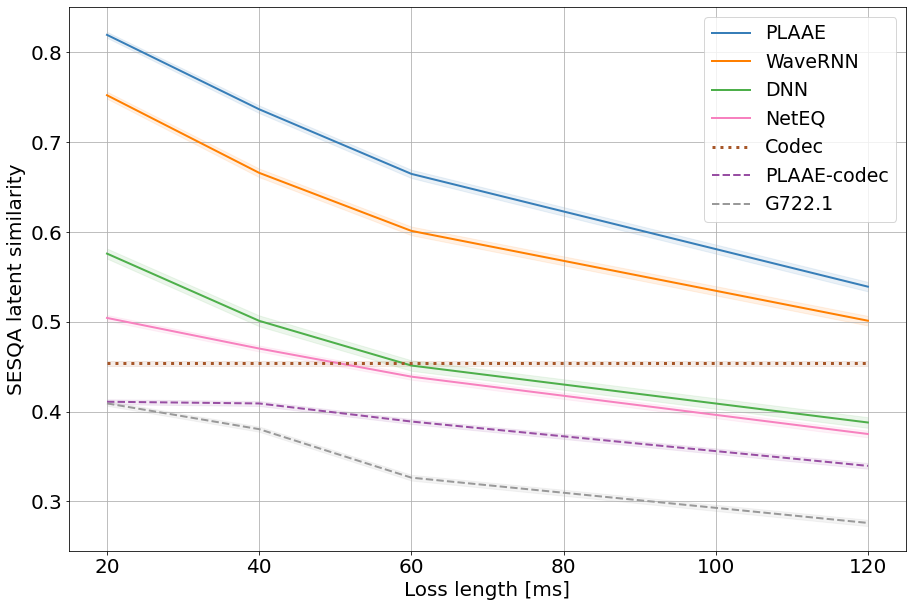}
    \vspace{-2mm}
    
    \caption{MCD (left) and SESQA latent similarity (right) as a function of packet loss lengths. Shaded regions show 95\% confidence intervals.}
    \label{fig:results}
\end{figure*}

The left plot of Fig.~\ref{fig:results} shows the MCD results corresponding to the signal distortion evaluation. Firstly, we can see that the distortion trend across models is increasing with gap length, as expected. Secondly, we see that PLAAE obtains a consistently lower distortion across all gap lengths, potentially due to its multi-STFT loss that directly optimizes the spectral similarity. This holds for both codec data (PLAAE-codec) and PCM data (PLAAE). On the other hand, the silence line poses a reference for the trivial zero filling baseline, and we can see that many models quickly converge towards that line after 40\,ms gaps. This is the expected behavior for classic models (NetEQ and G722.1), which typically fade-out to silence after a couple of packets, and some autoregressive models(DNN), which suffer from quick energy decay. On the other hand, the WaveRNN behaves erratically with this metric due to its generative nature on creating an alternative signal that sounds plausible but differs quickly from the real one. Table~\ref{tab:f0} contains the F0 and UV errors for the considered models. We see that PLAAE also attains the least error on both metrics, implying that it continues content and intonation more coherently. Note that the DNN yields a high error, contrary to the MCD result of Fig.~\ref{fig:results}. This is due to the inefficacy of training the phase reconstruction via regression, resulting into a noisy phase that provokes unvoiced sounding. 


\begin{table}[t]
\centering
\caption{F0 and UV error results, including 95\% confidence intervals. For both metrics, lower is better.}
\label{tab:f0}
\vspace{2mm}
\setlength{\tabcolsep}{10pt}
\begin{tabular}{l|c|c}
    \hline 
         Model   & F0 [Hz] & UV \\
     \hline
     Silence & 129.90 $\pm$ 3.95 & 0.49 $\pm$ 0.02 \\
     G722.1 & ~~59.30 $\pm$ 3.90 & 0.42 $\pm$ 0.02 \\
     WaveRNN  & ~~44.00 $\pm$ 3.10 & 0.36 $\pm$ 0.02 \\ 
     DNN & 100.33 $\pm$ 3.94 & 0.49 $\pm$ 0.02 \\
     NetEQ & ~~58.42 $\pm$ 3.79 & 0.36 $\pm$ 0.02 \\
     \hline
     PLAAE-codec (Ours) & ~~35.90 $\pm$ 3.12 & 0.28 $\pm$ 0.01 \\
     PLAAE (Ours) & \textbf{~~35.70} $\pm$ 3.10 & \textbf{0.27} $\pm$ 0.01\\
     \hline
\end{tabular}
\end{table}

Perceptual results from SESQA are depicted on the right side of Fig.~\ref{fig:results}. In this case, we can see that PLAAE attains the highest score across packet loss lengths, even when it may contain several losses in the recent past captured by its receptive field. We also observe that WaveRNN follows PLAAE in quality, contrary to what happened with the spectral envelope matching of the MCD score. As this metric evaluates the signal naturalness, WaveRNN scores high in similarity with respect to the real signal, as both follow the natural speech distribution. Nonetheless, the WaveRNN may be incoherent in the generated content due to its generative sampling, which causes the MCD curve to increase rapidly. Then, SESQA rates DNN and NetEQ as the next models with higher naturalness. Finally, we encounter the codec data and the models processing this input. 
In this codec regime, PLAAE still surpasses the naturalness of the G722.1 PLC algorithm.

Finally, intelligibility results are shown in Table~\ref{tab:cer} 
(models that process codec data are not evaluated in this scenario due to the ASR being pre-trained without those conditions). In this case, PLAAE also shows better longer-term resiliency than its counterparts. In the 20\,ms reconstruction case, it even shows a better CER than the lossless signal itself, obtaining a 6\% of relative improvement. We hypothesize that this is caused by PLAAE reconstructions reducing the variability of each phonetic unit's signal representation. Note that it guesses the content by context and recreates a representative of that content, which may contain less noisy nuances than the original signal. Moreover, the WaveRNN shows a rapidly increasing CER with loss length increase. This is aligned with the MCD increase of Fig.~\ref{fig:results}, and indicates that the sampling process is effectively degenerating into predicting content that is far from the expected one. We did our best to tune the $\softmax$ sampling temperature to obtain more stationary outcomes, but these very easily resulted in quickly energy-decaying signals, similar to those from the DNN. 

\begin{table}[t]
\centering
\caption{CER (\%) for different packet loss lengths. The CER for the lossless signal is 11.4\%. }
\label{tab:cer}
\vspace{2mm}
\begin{tabular}{l|c|c|c|c}
    \hline 
            Model & 20\,ms & 40\,ms & 60\,ms & 120\,ms \\
     \hline
     Silence & 17.3 & 26.3 & 35.1 & 49.0 \\
     WaveRNN  & 13.2 & 24.9 & 36.7 & 53.2 \\ 
     DNN & 13.6 & 22.5 & 31.4 & 45.1 \\
     NetEQ & 11.7 & 17.6 & 25.4 & 43.4 \\
     \hline
     PLAAE (Ours) & \textbf{10.7} & \textbf{14.1} & \textbf{20.0} & \textbf{37.8}  \\
     \hline
\end{tabular}
\end{table}

\section{Conclusion}
\label{sec:conclusions}
In this work, we propose the packet loss adversarial auto-encoder, or PLAAE. This model consumes a signal that is zero-filled in the lossy sections and learns to reconstruct them via an auto-encoding process. Hence, its inference process is parallelized, which contrasts with existing autoregressive approaches. This makes PLAAE suitable for real-time PLC applications by construction. The different proposed evaluation metrics, which assess the reconstruction quality, identity, and content matching, show that PLAAE consistently outperforms both existing classic and deep learning-based approaches through different loss lengths up to 120\,ms. 

\section{ACKNOWLEDGMENTS}
\label{sec:ack}

We would like to thank Giulio Cengarle, Paul Holmberg, Jiaquan Huo, and Tim Neal for fruitful discussions about PLC details.

\bibliographystyle{IEEEtran}
\bibliography{refs21}
%
%
%
%
%
%
%
%
%

 \end{sloppy}
\end{document}